\documentclass[a4,12pt]{article}

\usepackage{graphicx}

\usepackage{authblk}

\usepackage{algorithm}

\usepackage{algorithmic}

\usepackage[figuresright]{rotating} 

\usepackage{epsfig,times,latexsym,enumerate,lscape,flafter,amsmath,amssymb,graphicx}
\usepackage{blindtext}
\usepackage{graphicx}
\usepackage{adjustbox}
\usepackage{cite}
\usepackage{amsmath}
\usepackage{amsfonts}
\usepackage{lipsum}
\usepackage[bookmarks=false]{hyperref}

\begin{document}

\title{Comparison of Deep Learning and the Classical Machine Learning Algorithm for the Malware Detection}

\author{Mohit Sewak, Sanjay K. Sahay\thanks{E-mail: ssahay@bits-goa.ac.in} \hspace{0.01cm} and Hemant Rathore\\ 
Department of Computer Science and Information System, Birla Institute
of Technology and Science, K. K. Birla Goa Campus, NH-17B, By Pass Road, Zuarinagar-403726, Goa, India
}
\date{}

\maketitle

\begin{abstract}

Recently, Deep Learning has been showing promising results in various Artificial Intelligence applications like image recognition, natural language processing, language modeling, neural machine translation, etc. Although, in general, it is computationally more expensive as compared to classical machine learning techniques, their results are found to be more effective in some cases. Therefore, in this paper, we investigated and compared one of the Deep Learning Architecture called Deep Neural Network (DNN) with the classical Random Forest (RF) machine learning algorithm for the malware classification. We studied the performance of the classical RF and DNN with 2, 4 \& 7 layers architectures with the four different feature sets, and found that  irrespective of the features inputs, the classical RF accuracy outperforms the DNN.
\end{abstract}

\maketitle

\section{Introduction}
In today's world, information is one of the most valuable assets, but there is a major threat to it by the evolving sophisticated malware ({\bf mal}icious soft{\bf ware}). The first malware was created for fun, but now its a profit-driven industry \cite{security-software-statistics}  and in the last couple of years advanced malware uses complex obfuscation methods viz. control/data flow permutation, compression, heap spray, etc. \cite{You-2010} to evade the detection techniques. Also, there is an exponential increase in the number of malware released every year. To detect these malware, various techniques are proposed in the past. These methods range from the early day signature-based detection to Machine Learning techniques\cite{Ashu-Survey}. However, it appears that proposed approaches are not sufficient to detect the unknown sophisticated malware. Thus researchers are continuously searching for better techniques to detect the advanced malware and have started applying machine learning techniques for malware detection. These algorithms can detect obfuscated malware and have potential to handle both complexity and scale problem like data pre-processing, feature extraction/selection, model creation and testing. In this  Santos et. al claims to obtain an accuracy of $\sim$96\% using opcode frequency as features \cite{Santos-2013}. Later on, researchers also applied file-size based segmentation on opcode frequency and improved the average accuracy using different Machine Learning models and claimed that Random Forest can provide the accuracy up to $\sim$98\% with a False Positive Rate (FPR) of $\sim$1.07\% \cite{Ashu-RF}. For the analysis they used the benchmark malware samples from the Malicia project \cite{Malicia}.

Recently, Deep Learning has been showing promising results in various Artificial Intelligence applications like image recognition, natural language processing, language modeling, neural machine translation, etc. \cite{sewak-cnn}. Therefore, in this paper, we compared the effectiveness of a Deep Learning Architecture called Deep Neural Network (DNN) with the classical Random Forest (RF) machine learning algorithm for the malware classification. We did not use the segmentation method as used in the above work. Instead, we used different feature selection/extraction techniques and achieved a much better accuracy ($\sim$99\%) using Random Forest with an even better FPR ($\sim$0.24\%). Our analysis indicates that Deep Learning based architectures such as Auto-Encoders for feature extraction combining with Deep Neural Networks for the classification does not give statistically  better accuracy  as compared to state-of-art machine learning technique combinations like Variance Threshold (for feature selection) and Random Forest (for classification) on Malicia data set. Rest of the article is organized as follows. Section 2, in brief, discusses the related work done in the malware classification. Section 3 explains the data preprocessing, feature extraction and selection approach for comparing the model. Section 4 compares the results of DNN (2, 4  and 7 layers) with classical RF classifier and Section 5 finally contains the conclusion of the article.

\section{Related Work}

There are two basic approaches to identify malware from the benign programs, namely static analysis, and dynamic analysis.
In the static approach without executing the program  features  are extracted from the file or its associated metadata, and then using the selected/extracted features (e.g., opcode frequency \cite{ Ashu-RF}, strings \& byte sequence \cite{ Schultz-2001}, function length \cite{ Tian-2008}, API Calls \cite{Xin-2016}, etc.) the detection algorithm is applied for the classification. Whereas in the dynamic analysis, the program is executed, and then important behavioral aspects and other parameter are captured as features on which the detection algorithm is applied. In addition, there also exists approaches where different features from both the static and dynamic analysis are combined to form a Hybrid approach \cite{Ahmadi-2016}. Bilar \cite{Bilar-2007} directly used op-code frequency distribution for malware detection whereas Karim et al. \cite{Karim-2005} used op-code sequences and permutations for the same. Ashu et al. applied file-size based grouping and opcode frequency as features to improved the average accuracy with different Machine Learning models with Malicia data set \cite{Malicia}, and obtained accuracy up to 98\%.

For the classification of malware by the supervised machine learning, typically a binary model is developed to differentiate malware and benign programs, and sometimes multinomial models are also constructed for distinguishing samples across the different malware families. In this some popular supervised machine learning algorithm used for malware classification are Naive Bayes \cite{ Kolter-2004}, Support Vector Machine \cite{Kolter-2004, Santos-2013}, Decision tree \cite{Kolter-2004}, Iterative Dichotomiser 3, J48 \cite{Ashu-RF}, Hidden Markov Models, Random Forest \cite{Ashu-RF}, AdaBoost, Instance-Based Learning Algorithms \cite{ Kolter-2004} and TF-IDF \cite{Kolter-2004, Santos-2013}. Also, for feature extraction/selection/dimension reduction algorithms can be also used with supervised learning to boost its performance. The popular way of feature extraction is to use an unsupervised learning algorithm for making better representative features at the supervised learning stage, e.g., feature selection methods such as hierarchical, unary variable removal, Goodness evaluator, and Weighted Term Frequency \cite{Santos-2013}. However, if feature extraction is not done properly, then it may result in lesser accuracy, higher false positive rates, etc. Due to these shortcomings, Deep Learning based techniques are gaining a lot of traction for the feature extraction and otherwise also. Recently a lot of efforts has been focused and  achieved success with different Deep Learning techniques like Deep Belief Networks \cite{DeepSign-2015}, Deep Neural Networks \cite {Saxe-2015}, Recurrent Neural Networks (RNN) (its variants like Long Short-Term Memory (LSTM) \& Gated Recurrent Unit (GRU)) \cite{Pascanu-2015} and combination of Recurrent and Convolutional Neural Networks (CNN) \cite{Kolosnjaji-2016} for supervised learning and classification. Similarly, for optimal feature extraction Deep Belief Networks (DBN) and (Stacked) Auto-Encoders (SAE/ AE), and RNN based Auto-Encoders (RNN-AE) \cite {Xin-2016} have been used.

\section{Data Pre-Processing and Feature Selection/ Extraction}

For the investigation of the popular Random Forest and Deep Neural Networks with 2, 4, 7 layers, we use the 11,308 malicious files downloaded from benchmark Malicia project (one of author posses the dataset \cite{Ashu-RF}) with different sets of features for the classification of malware. For the analysis, we collected 2,819 benign files (cross-verified from the `virustotal.com' \cite{VirusTotal} file scanning service) from different Windows systems. As the number of benign and malicious files are not comparable, we used Adaptive Synthetic (ADASYN)\cite{ADASYN} technique to alleviate the class imbalance problem. Now using the Linux \textit{objdump} utility, we extracted the opcodes of all the executables and made a cumulative list of all the unique opcodes (\textit{master opcode list}). The extracted unique opcodes are mapped with an integer for further analysis. Then we computed the frequency of each opcode of every executable, and ordered them as per the master opcode list for the comparison of the classifiers under investigation.

\begin{figure}
    \centering
    \includegraphics[width=3.4in,height=6in]{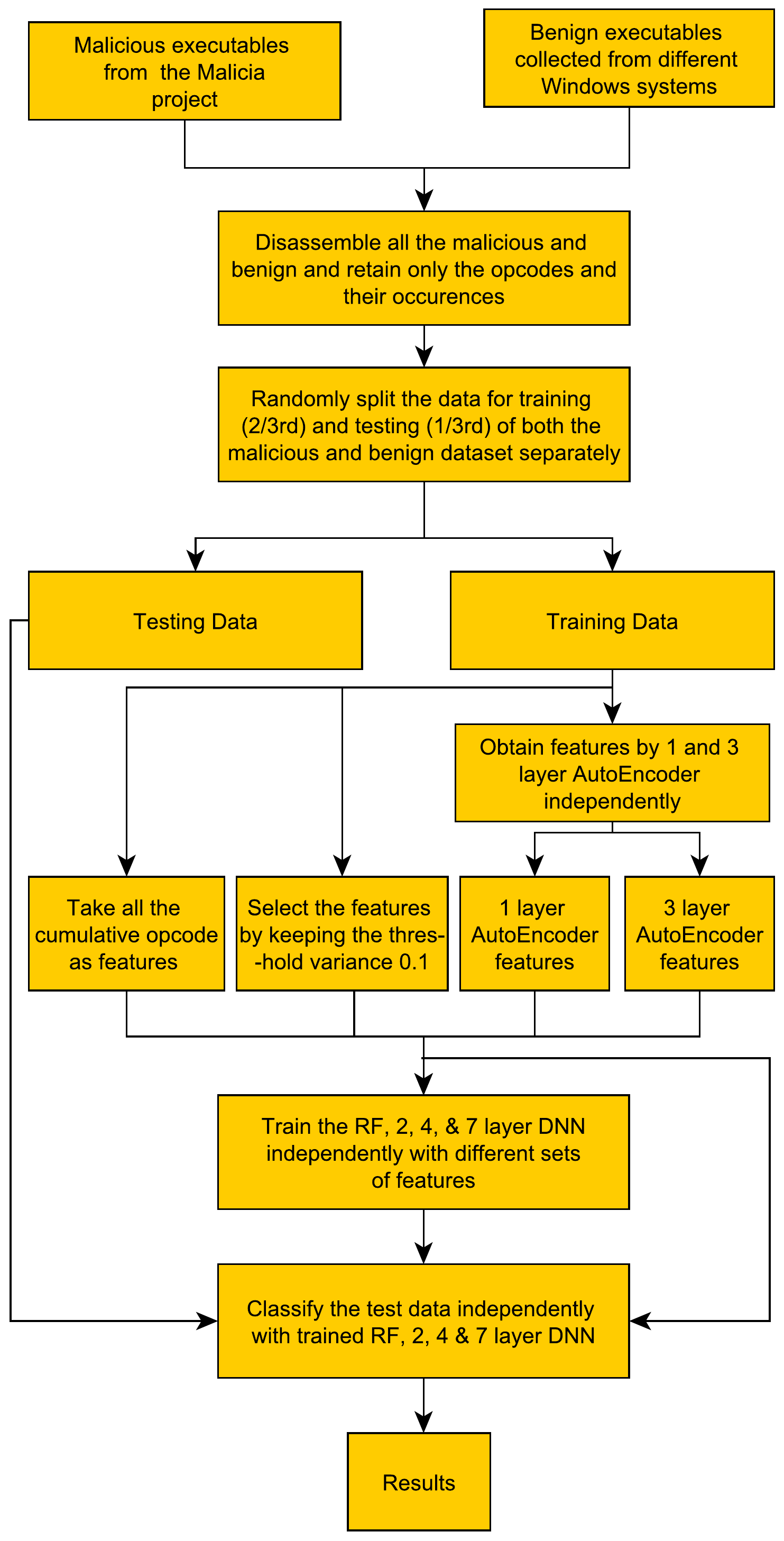}
    \caption{Flowchart for the classification of malware by RF, 2, 4, 7 layers DNN with different sets of features.}
    \label{figure:Flowchart}
\end{figure}

The flowchart \ref{figure:Flowchart} shows our approach to classify the malware using Random Forest, 2, 4, and 7 layers Deep Neural Network with the different sets of feature viz. by

\begin{enumerate}
    \item taking all the cumulative opcodes, i.e., all the 1613 opcodes occurrence has been used for the classification.
    \item selecting the features by keeping the variance threshold (VT) to 0.1, in this case, 616 opcodes occurrence has been used for the classification. 
    \item extracting the features by AutoEncoder with one layer (AE 1L) and three layers (AE 3L). In this input layer takes all the 1613 opcodes occurrence, and output layer has been set to 32 for both AE 1L and AE 3L. 
\end{enumerate}
\medskip

Now we randomly split the dataset (containing only opcode occurrence of every executable) for the training and testing of the malicious and benign dataset separately (a conservative side as per the suggested norms to ensure optimal performance \cite{Guyon-1997}) in the ratio of 2:1.

\section{Comparison of Deep Leaning and an effective Machine Learning Random Forest algorithm for the Malware detection}

To compare the most popular and effective Machine Learning Random Forest classifier with Deep Learning approaches, we had chosen Deep Neural Network with 2, 4, and 7 layers. For the Random Forest, we use the Python library Sci-Kit Learn \cite{sk-learn}, for DNN and AE we use the Python library Keras \cite{keras} with TensorFlow \cite{tensorflow} back-end. In DNN, three different layers (2, 4 and 7) architectures have been chosen to understand the performance of the DNN with the increase in the number of hidden layers. Here all the DNN layers use Exponential Linear Units (ELU) \cite{ELU} as activation function, except for the last layer, i.e., prediction layer uses Sigmoid activation function, and the model are trained by setting the dropout rate to 0.1 with a batch size of 64. Also, DNN uses Adam \cite{adam-grad} optimizer and binary cross-entropy ($H(p,q)$) \cite{cross_entopy} loss function for the distribution $p$ (actual class) and $q$ (probability of the predicted class) over a set of events ($x$) given as:

$$ H(p,q) = - \sum_x p(x) \log q(x) $$

To compare the performance of the Random Forest with 100 trees in the forest, 2, 4, 7 layers DNN, we computed the True Positive Rate (TPR) / Sensitivity / Recall / Hit Rate, True Negative Rate (TNR) / Specificity, Positive Predictive Value (PPV) / Precision and Accuracy (Acc.) defined as \cite{evaluation_measure}:

{\centering {$$ \mbox{TPR} = \frac{TP}{TM}; \quad \mbox{TNR} = \frac{TN}{TB}; \quad \mbox{PPV} = \frac{TP}{TP + FP}$$}
$$ \mbox{Accuracy} = \frac{TP + TN}{TM + TB}$$}
\noindent where,
\begin{table}[h]
\centering
\normalsize
\begin{tabular}{ccl}
TP & $\longrightarrow$ &True positives; the number of malware instances\\
&& correctly classified. \\
\indent TN & $\longrightarrow$ &  True negatives; the number of benign instances \\
&&correctly classified. \\
\indent FP &  $\longrightarrow$  & False positives; the number of benign instances \\
&&wrongly detected as malware. \\
\indent TM  & $\longrightarrow$  & Total number of malware instances. \\
\indent TB &  $\longrightarrow$  & True number of benign instances. \\
\end{tabular}
\end{table}

From the investigation (Table \ref{table:result}), we found that overall the Random Forest's results are better than the 2, 4, 7 layers DNN for all the four different sets of features viz. None (i.e., all the cumulative opcodes taken as the feature), VT, AE-1L, and AE-3L. However, if compared between all the DNN, seven layers DNN performance is the best (99.21\% accuracy).

In the classification, features inputs to the classifier play a vital role. Therefore in terms of input features the analysis shows that classical RF performance is more or less same with None or VT features. DNN-2L gives the best accuracy with the features obtained by VT, whereas DNN with 4 layers gives the highest accuracy with AE-1L. However, among the DNN's (with different features inputs) AE-1L provides the maximum accuracy (99.21\%) but if the feature input is from AE-3L, then there is $\sim$6\% decrease in the classification accuracy.

\begin{table}
\centering
\begin{tabular}{|c|c|c|c|c|c|}
\hline
Classifiers &  Features & Acc. & TPR & TNR & PPV \\
\hline
{\bf RF} &  None & {\bf 0.9974} & 0.9948 & 1.0000 & 1.0000 \\
DNN 2L & None & 0.9779 & 0.9633 & 0.9926 & 0.9924  \\
DNN 4L &  None& 0.9742 & 0.9538 & 0.9948 & 0.9946  \\
DNN 7L &  None & 0.9615 & 0.9905 & 0.932 & 0.9366  \\
\hline
{\bf RF}  &  VT & {\bf 0.9978} & 0.9959 & 0.9997 & 0.9997  \\
DNN 2L & VT & 0.9884 & 0.9832 & 0.9937 & 0.9937  \\
DNN 4L &  VT& 0.9869 & 0.9796 & 0.9942 & 0.9942  \\
DNN 7L &  VT& 0.9620 & 0.9889 & 0.9348 & 0.9389  \\
\hline
{\bf RF}  &  AE 1L & {\bf 0.9941} & 0.9886 & 0.9997 & 0.9997  \\
DNN 2L &  AE 1L & 0.9695 & 0.9457 & 0.9937 & 0.9934  \\
DNN 4L & AE 1L & 0.9899 & 0.9829 & 0.9970 & 0.9970  \\
DNN 7L & AE 1L & 0.9921 & 0.9861 & 0.9981 & 0.9981  \\
\hline
{\bf RF} & AE 3L  & {\bf 0.9936} & 0.9872 & 1.0000 & 1.0000  \\
DNN 2L & AE 3L & 0.9625 & 0.9375 & 0.9879 & 0.9874  \\
DNN 4L & AE 3L & 0.9716 & 0.9861 & 0.9568 & 0.9585  \\
DNN 7L & AE 3L & 0.9360 & 0.8797 & 0.9931 & 0.9923 \\
\hline
\end{tabular}

\medskip

\caption{Results with multiple feature methods and classification models}
\label{table:result}
\end{table} 


\section{Conclusion}
In this paper, we compare the performance of the classical machine learning  Random Forest techniques and Deep Neural Networks (with multiple architectures) for the classification of malware with different sets of features. We found that overall classical ML algorithm Random Forest outperform the different layers architecture of DNN. The best accuracy obtained by RF and DNN are 99.78\% and 99.21\% with feature set obtained by VT and AE-1L respectively. This indicates that Deep Learning based architectures such as Auto-Encoders (used for feature extraction) and when combined with Deep Neural Networks (for the classification) may be an overkill the Malicia Dataset because its obfuscation technique may not be too complex to predict the malware using opcode frequency as a feature. Therefore it does not provide statistically better accuracy as compared to classical machine learning RF. However, further investigation is required to understand the effectiveness of Deep Learning techniques like RNN, LSTM, ESN, etc. coupled with more advanced data processing and feature extraction methods, and the work in this direction is in progress.

\bibliographystyle{unsrt}
\bibliography{bib_Malware_DL}

\end{document}